\documentclass[twocolumn]{IEEEtran}
\IEEEoverridecommandlockouts
\usepackage{cite}
\usepackage{amsmath,amssymb,amsfonts}
\usepackage{algorithmic}
\usepackage{graphicx}
\usepackage{textcomp}
\usepackage{xcolor}
\def\BibTeX{{\rm B\kern-.05em{\sc i\kern-.025em b}\kern-.08em
    T\kern-.1667em\lower.7ex\hbox{E}\kern-.125emX}}
\begin{document}

\title{Augmented Reality, Cyber-Physical Systems, and Feedback Control for Additive Manufacturing:\\ A Review
\thanks{$^*$ Joint first authors.}
\thanks{This publication has emanated from research supported in part by a research grant from Science Foundation Ireland (SFI) under grant number 16/RC/3872 and is co-funded under the European Regional Development Fund and by I-Form industry partners.}
}

\author{\IEEEauthorblockN{Hugo Lhachemi$^*$, Ammar Malik$^*$, Robert Shorten}\\
\IEEEauthorblockA{\textit{School of Electrical and Electronic Engineering} \\
\textit{University College Dublin}\\
Dublin 4, Ireland \\
hugo.lhachemi@ucd.ie, ammar.malik@ucdconnect.ie, robert.shorten@ucd.ie}
}

\maketitle

\begin{abstract}
Our objective in this paper is to review the application of feedback ideas in the area of additive manufacturing. Both the application of feedback control to the 3D printing process, and the application of feedback theory to enable users to interact better with machines, are reviewed. Where appropriate, opportunities for future work are highlighted.
\end{abstract}


\section{Additive manufacturing: a disruptive technology}

Additive manufacturing (AM), commonly known as {3D} printing, refers to the various processes (power bed fusion, direct deposit) of adding together materials, based on a {3D} model files for producing three-dimensional objects. The early developments of modern AM processes can be traced back to the 50s~\cite{RN113,RN114}, with the first  commercial application of AM emerging in 1987. This first commercial work pioneered the technique of stereolithography, in which a layer-by-layer {3D} printing technology was developed using photopolymerization~\cite{RN115}. Originally used for fast prototyping~\cite{RN41,RN82,RN83}, and as a visualization tool, AM has seen rapid growth during the last decade due to the advancement in processes, tools, and applications for both end-user part production and manufacturing at a large scale. AM is now on the cusp of widespread adoption in both homes and workplaces. It offers possibilities to not only print bespoke products but also the possibility of enabling highly disruptive new business models that are based around prosumers. Driven by these new possibilities, the research and industry communities are focused on not only making {3D} printing machines work better, but also on developing new ways  to enable humans to interact with such machines, in a feedback loop, and to enable networks of machines to interact with each other. This fast-growing research area, badged loosely as Industry 4.0, brings together ideas from machine learning, augmented reality, human-computer interaction, as well as the physics and chemistry of {3D} printing machines, all with the objective of solving human-in-the-loop control problems. As such it offers great opportunities for researchers to discover and solve new types of feedback control problems. It is in this context that we have written this paper; namely, to bring to the attention of the wider control community, the types of problems and opportunities that exist in this emerging area. 

This paper is structured as follows. The breakthroughs introduced by {3D} printing in the context of manufacturing, as well as their potential impacts in the emergence of a prosumer society, are described in Section~\ref{sec: primer on manufacturing}. Two of the main challenges faced by additive manufacturing that fall under the area of expertise of the control community are stated in Section~\ref{sec: contributions} and then investigated in the next sections. First, the importance of the low-level control strategy for additive manufacturing, the underlying challenges, and the related current state of the art are presented in Section~\ref{sec: low-level control}. Then, Section~\ref{sec: creation and iteration with 3D models} describes the opportunities for the use of augmented reality in the context of 3D modeling and human-machine interaction for additive manufacturing. Finally, concluding remarks are provided in Section~\ref{sec: perspectives and conclusion}.

\section{Primer on manufacturing}\label{sec: primer on manufacturing}

It is no understatement to suggest that modern societies have been shaped by the advances in manufacturing. It is not inconceivable that the rise of the prosumer society will similarly act as a catalyst for change in the coming decades, and will enable entirely new consumption habits. We begin our paper with a brief overview of manufacturing and the transformation in consumer consumption habits.  

\subsection{Transformation of the manufacturing processes}
Traditional (or conventional) manufacturing refers to processes that  consist of machining a part from a workpiece by removing material. This is also called subtractive manufacturing. In these types of manufacturing processes material is removed by both direct contact and relative motion of the cutting tool and the workpiece, and energy is used to rotate either the tool or the workpiece. In traditional manufacturing, the tooling environment (including the cutting tools, but also the patterns, the molds and the fixtures) is usually the most costly and time-consuming part of the process. Consequently, unit costs are very high for small-batch productions but drop considerably for mass productions; and these costs increase with the complexity of the part to be manufactured. 

Given this background, AM has emerged as a disruptive technology poised to deeply transform manufacturing: ``Additive manufacturing technology can break existing performance trade-offs by reducing the capital needed to achieve scale and scope economies.''~\cite{RN118}. First, AM obviates the need for dedicated tooling. Indeed, while traditional production lines require major tooling adaptations for producing different products, AM equipment remains mostly unchanged as different parts are produced. Furthermore, AM intrinsically reduces the unit cost of production when compared to subtractive manufacturing methods by reducing the quantity of material required for production, and by being, for the  most part, insensitive to the complexity of the part~\cite{RN117}. Finally, AM offers the capability to manufacture shapes that are not manufacturable with traditional manufacturing. Such a higher degree of flexibility can be used, for example, to manufacture lightweight parts that require less assembly time, products with customized features, and multi-material printed parts exhibiting space varying properties~\cite{RN116}.

\subsection{Transformation of the consumption habits}
Changes in manufacturing have acted as a catalyst for profound societal change. In the late 18th century, the first industrial revolution was characterised by the advent of factories using machines taking advantage of steam power as replacement for hand production methods. The second industrial revolution, which spanned the late 19th and the beginning of the 20th centuries, was characterised by an unprecedented wave of innovations, driven by electric power and the emergence of production lines allowing mass production. In the early 1970s, the emergence of electronics, computation units, and information technologies enabled the third industrial revolution through the massive automation of the manufacturing processes. Modern societies are now entering in a fourth industrial revolution, driven by ubiquitous connectivity and the internet of things. Sometimes called Industry 4.0, this new stage of manufacturing development is characterised by the emergence of smart factories and smart services providers developing smart products and smart services. 

The first three industrial revolutions led to standardized and uniformed products. As famously said by Ford in 1909 about model T: "Any customer can have a car painted any color that he wants so long as it is black"~\cite{RN94}. Customers are now no longer satisfied with this one-size-fits-all approach to manufacturing and are increasingly demanding bespoke products. 
This is driven, not only by consumer demands for personalized and customized products, but also by the needs of certain industries (e.g., in the context of health with prosthesis, dental implants, etc) which fundamentally require bespoke manufacturing. All these trends require not only advances in manufacturing, but also a bridging of the gap between the consumer, the designer, and the manufacturing process.

In this context, AM has emerged as a disruptive technology poised to deeply transform manufacturing~\cite{RN6,RN31}. Conventional factories relying on classic manufacturing methods take advantage of highly specialized production lines for producing standardized products. This approach was motivated by the paradigm: the more units produced, the lower the cost per unit. The first three industrial revolutions were driven by this paradigm, allowing access to mass consumption at an affordable cost by producing a great volume of units of a given good. On the other hand, highly specialized production lines are incompatible with the demand of modern consumers for specialized and personalized products as their adaptation to the production of very specific products in a small volume of units is too costly. The big breakthrough resulting from additive manufacturing is the fact that the production cost per unit is mostly insensitive to the number of units produced~\cite{RN31}. Therefore, while it is always more cost effective to resort to conventional manufacturing methods for mass productions, it is generally quicker and much more affordable to resort to additive manufacturing for small-batch productions of highly personalized products. 

{3D} printing also narrows the gap between the consumer, the designer, and the production. While traditional manufacturing was the reserve of factories, Fused Deposition Modeling (FDM) AM processes~\cite{RN159,RN160} allow the transfer of the production capabilities from factories to small companies and homes taking advantage of: 
\begin{itemize}
\item the wide use of the personal computers and of the internet;
\item the easy access to tutorials, user-friendly software, and banks of {3D} models;
\item rapid progresses in printing tools, quality, and speed;
\item the ever-reducing costs of the FDM {3D} printers and related printing materials (e.g., thermoplastic filaments).
\end{itemize}
Therefore, while the consumer, the designer, and the manufacturer were three distinct agents in traditional manufacturing, {3D} printing allows merges the three roles into a single one. This is sometimes referred to as the emergence of the prosumer~\cite{RN229}.

\section{Contributions of the paper}\label{sec: contributions}
Our objective in this paper is to discuss the application of feedback to make {3D} printing better. We focus on two main areas.

\subsection*{Feedback based on traditional control techniques}
The first objective of this paper is to discuss classical feedback control in the design of {3D} printing systems. For the most part, the discussion of classical feedback theory is restricted to the operation of the {3D} printing process. Where appropriate, we should also discuss the application of such theory in other parts of the {3D} printing process.

\subsection*{Feedback based on augmented reality}
The use of augmented reality as a feedback tool is limitless. Much of the present paper is devoted to describing the many existing applications to better the manufacturing process. Roughly speaking, AR enables the augmentation of the user's real world with immersive computer-generated information such as visuals, sounds, touch interactions, etc~\cite{RN74}. Although the first basic AR prototypes were developed in the 1960s~\cite{RN95}, AR has emerged over the past two decades as one of the most exciting fields in many application domains. Presently, a great number of AR technologies are used. For example, the interested readers are referred to~\cite{RN74,RN96,RN99,RN100,RN97,RN98,RN102,RN104,RN101,RN106,RN108,RN105,RN107}. Currently, design and manufacturing represent two of the most important applications of AR~\cite{RN78,RN110}. The main reason is that AR offers the opportunity for designers and operators to interact in an intuitive manner, directly in their physical space, with the information of the creation/production process. For instance, CAD tools can be merged with virtual reality or AR-based technologies, providing immersive modeling environments~\cite{RN111,RN112} or interactive virtual assembling environments~\cite{RN109}. The first objective of this paper is to provide a comprehensive review of the opportunities offered by AR for {3D} manufacturing.

\section{Low-level control}\label{sec: low-level control}

In spite of its huge potential~\cite{RN116,RN118}, AM still does not match the standards of conventional manufacturing. Typically, AM processes are characterised by low productivity, poor quality, inconsistent reproduction properties, and uncertain properties of the manufactured parts ~\cite{RN142}. These pitfalls prevent the widespread adoption of AM technologies in industrial sectors with stringent precision requirements such as aerospace~\cite{RN135} and biomedical~\cite{RN136} industries. The essential cause of these imprecisions is the inherent difficulty to model, monitor, and control the underlying AM processes~\cite{RN142}. Furthermore, AM covers a large broad of technologies and processes requiring specific and widely differing modeling and control strategies. A complete review mapping AM modeling approaches is available in~\cite{RN134}. In this section, we review the low-level control strategies reported for two of the most promising AM technologies; namely, extrusion-based {3D} printing and the metal-based AM.

\subsection{Extrusion-based {3D} printing}
Extrusion-based {3D} printing technologies are among the most promising technologies for fast prototyping, small batch productions of plastic parts, and for making manufacturing accessible  to small companies and homes. This type of AM covers a number of technologies such as FDM~\cite{RN159,RN160}, syringe-based extrusion~\cite{RN162}, and screw extrusion~\cite{RN161,RN177}. These essentially consist of the heating, and the pressurization of plastic materials in an extruder chamber before deposition in the printing workspace.

Apart from the control of the extruder position~\cite{RN165}, the essential aspects of extrusion-based {3D} printing processes that require feedback control are the extruder melt temperature and the extrusion flow rate. One essential aspect, referred to extrusion-on-demand~\cite{RN225}, relies on the precise control of the extrusion flow rate for achieving a delivery {\em on-demand} of the material. Such an extrusion-on-demand is critical for the printing of complex geometries which generally require discontinuities in the material delivery. Furthermore, in order to achieve high quality of the extruded product, the control laws must be robust against certain external disturbances~\cite{RN166}. These include any blocks of solid polymer that are not sufficiently melted, air bubble induced compressible behavior of the paste, as well as pressure disturbances and non-homogeneity in the extrusion flow, etc. Both the value and the homogeneity of the extruder melt temperature are among the most important process parameters. Indeed, it is known that the failure to ensure adequate thermal conditions can lead, for example, to thermal degradation, inadequate mechanical properties of the final part, geometrical inaccuracy, and non-homogeneity in the extrusion flow rate \cite{RN168,RN173}. Furthermore, the ability to ensure the homogeneity of the melt temperature relies on numerous control parameters including the processing conditions and material properties~\cite{RN169,RN170}. Most of the control strategies for extrusion reported in the literature to control the melt temperature employ a linear model of the plant (to design a simple proportional integral derivative (PID) controller)~\cite{RN171,RN172}. For example, a PID controller with anti-windup strategy has been reported in~\cite{RN178}. A multi-objective model predictive control strategy has also been proposed in~\cite{RN173} allowing the inclusion of safety requirements in the control design. In~\cite{RN182}, a three-stage control law has been proposed to control the volumetric flow of a single-screw extruder via joint regulation of both temperature and pressure. First, an inner-loop, composed of seven parallel PID controllers, are used to control the local temperatures along the barrel. Then, a PID-based outer-loop is employed to control the temperature at the output of the extruder. Finally, a PID controller is considered for regulating the output pressure. In this setting, a multivariable generalized predictive control law has been designed in ~\cite{RN188} along with a feedforward component that avoids sudden variations of temperatures when the pressure is changed. In order to compensate for changes in process conditions, adaptive controllers have been proposed, for example in~\cite{RN172,RN184}, while a disturbance decoupling control design has been investigated in~\cite{RN187}. Due to the nonlinear behaviour of the system, linear time-invariant controllers may fail to ensure the required {\em die} temperature set-point tracking performance over the full operating range of the system. In this context, a nonlinear dynamic model was developed in~\cite{RN167}. Based on the measurement of the die melt temperature via an infrared sensor, the proposed control strategy is based on fuzzy logic control. Fuzzy control strategies in extrusion-based {3D} printing were further investigated in~\cite{RN185}. Similarly, a PID controller with the online tuning of its gains by means of a neural network has been proposed in~\cite{RN186}. In this setting, the neural network is employed for both online identification of the plant, and tuning of the controller gains. PDE modeling along with nonlinear model predictive control strategies for a twin-screw extruder have been reported in~\cite{RN179,RN180}. Beyond the control of the die melt temperature and pressure, certain control strategies have also been reported for controlling other relevant physical quantities such as the die viscosity with a PID controller in~\cite{RN190} or by means of fuzzy control in~\cite{RN213,RN214}.

Studies have also been reported on the control of the extrudate geometrical characteristics~\cite{RN176}. In~\cite{RN174}, it has been proposed to control the extrudate thickness by means of a proportional integral (PI) controller and a Smith predictor. This study has been extended in~\cite{RN175} to the control of the extrudate shape and size. Two control algorithms have been compared, namely a simple PI controller and a multivariable feedforward plus feedback controller. In~\cite{RN183}, a multivariable adaptive and control scheme has been proposed for controlling both the temperature and the thickness of the extrudate. The possibility to directly control the final part geometry in order to avoid the manual adjustment of the process parameters by technicians has been investigated in~\cite{RN189}. The proposed control strategy relies on an iterative learning controller and its relevance was demonstrated for the regulation of the width of a rectangle.

The control of the start and stop of extrusion-on-demand remains challenging (change in paste properties, inhomogeneities, air bubble release). The regulation of the extrusion force was investigated in~\cite{RN226} by means of an adaptive controller that estimate in real-time the parameters of the plant. In~\cite{RN228}, a hierarchical control structure is developed for controlling the material delivery flow. The proposed approach combines the control of both extruder ram velocity and pressure applied to the paste. The control structure consists in a model predictive controller that is supervised by a decision algorithm that switches between extrusion force control and ram velocity control. Similarly, a hybrid extrusion force-velocity controller is designed in~\cite{RN163} to achieve an extrusion-on-demand with air bubble release compensation. Specifically, a first controller is designed based on a first order model of the syringe's flow rate in order to ensure a steady-state extrusion flow rate. However, plunger velocity control exhibits slow dynamics while fast response is required for either extrusion-on-demand or to compensate disturbances resulting from the release of air bubbles. For this reason, an extrusion force controller is designed based on the following first order dynamic extrusion force model~\cite{RN164}: 
\begin{equation*}
    \dfrac{\mathrm{d} f_L}{\mathrm{d}t}(t)
    = \dfrac{( f_L(t) - f_f + A_p p_a )^2}{A_p p_0 l_0} (u_p(t) - u_a(t)) ,
\end{equation*}
where $u_p$ is the plunger velocity, $f_L$ is the extrusion force, $f_f$ is the friction force between the paste and the barrel, $A_p$ is the cross-sectional area of the plunger, $P_a$ is the atmospheric pressure, $p_0$ is the initial pressure in the syringe, $l_0$ is an effective air layer thickness, and $u_a$ is the steady-state velocity corresponding to a given extrusion force. The hybrid extrusion force-velocity controller is obtained by means of a switching signal. Specifically, the extrusion force controller is used to precisely control the extrusion start and stop while the plunger velocity controller is used most of the build time to regulate the output of the syringes. One important aspect to be considered for improving the performance of the closed-loop system is the delay of extrusion start and stop, referred as the dwell-time in~\cite{RN227}. This issue is tackled in~\cite{RN191,RN192} by modeling the extrusion-on demand process via a 1D hyperbolic PDE coupled with an ordinary differential equation. Specifically, denoting by $L$ the length of the extruder and by $x(t)$ the moving boundary boundary delimiting the part of the extruder that is filled of material, the mass balance equation of the filled zone yields for an incompressible homogeneous material with constant density $\rho_0$ yields the following {1D} hyperbolic PDE:
\begin{align*}
    \dfrac{\partial u}{\partial t} (t,z)
    & = \xi N_0 \dfrac{\partial u}{\partial z} (t,z) ,
    & (t,\xi) \in \mathbb{R}_+ \times (x(t),L) \\
    u(L,t)
    & = \dfrac{F_\mathrm{in}(t)}{\rho_0 N_o V_\mathrm{eff}} ,
    & t \geq 0
\end{align*}
where $u(t,z)$ represents the distributed filling ratio, $F_\mathrm{in}(t)$ is the feed rate used as a control input, $N_0$ and $\xi$ denote the screw speed and pitch, respectively, and $V_\mathrm{eff}$ is the effective volume between a screw element and the extruder barrel. This PDE exhibits a time-varying domain $(x(t),L)$ whose dynamics is described by the following ODE:
\begin{equation*}
\dfrac{\mathrm{d} x}{\mathrm{d} t}(t)
= - \xi N_0 \dfrac{ K_d x(t) - ( B \rho_0 + K_d x(t) ) u(t,x(t)) }{(B \rho_0 + K_d x(t)) ( 1 - u(t,x(t)))}
\end{equation*}
where $B$ represents a screw geometric parameter while $K_d$ stands for the nozzle conductance. The control design of this model was investigated in~\cite{RN191,RN192} and consists in a state-dependent input delay-compensated bang-bang controller.

\subsection{Metal-based additive manufacturing}
The term metal-based AM gathers together the most promising AM technologies for industrial applications for areas such as aerospace~\cite{RN135}, automotive~\cite{RN116,RN137}, and biomedical~\cite{RN136} industries. Such methods are essentially composed of three main steps. First, the {3D} Computer Aided Design (CAD) model of the part is numerically processed to be sliced into layers. Then, a high power source of energy, typically an electron or a laser beam, is used to melt over the current layer the metal wire or powder (either powder-bed or powder-fed systems). By cooling down, the metal forms a dense layer which contributes to form the final {3D} printed part. Finally, the process is repeated over successive layers to print the full {3D} model. Even though metal-based AM has emerged as a disruptive technology for both the manufacturing and the repair of metallic parts, numerous challenges remain to bring it to the level of quality and repeatability reached by traditional manufacturing processes. Indeed, while AM offers many advantages~\cite{RN135,RN136} in comparison to traditional manufacturing processes, it is crucial that AM processes are accurately modeled to predict the material and mechanical properties of the manufactured part. However, complex interactions arise in there direct material deposition processes due to the interactions between the laser beam, the powder, and the substrate. Numerous parameters, including process parameters (laser power, powder feed rate, scan speed, trajectory generation, etc.), ambient parameters (temperature or humidity), and intrinsic properties of the employed materials have a strong impact on the quality of the final product. Any deviation of one of these parameters might result in defects. For instance, due to the off-line generation and optimization of the printing trajectories, the accuracy of the geometry might not reach the required level. Because of the layer-based process, inhomogeneities can appear yielding problems in mechanical and material properties. For direct metal deposition, a too deep/large melt pool can induce irregularities due to re-melt of the solidified material. On the other hand, a too thin melt pool does not allow the required high-strength bonding between the deposited layers~\cite{RN119}. In this context, there is a strong interest in the development of advanced physics-based models, sensing technologies~\cite{RN145,RN205}, monitoring~\cite{RN146}, and control strategies for AM~\cite{RN142}. We now give a brief review of the modeling and control approaches in this area.

\subsubsection{Modeling}
The understanding of the complex interactions between the process and the physical properties of the manufactured part, including build conditions, micro-structures, residual stress, thermal distortions, etc~\cite{RN119}, is an active research topic. A complete review of metal-based AM processes, including their impact on the structure and properties of the fabricated metallic components, can be found in~\cite{RN121,RN143}. Based on the understanding of the underlying phenomenon, the development of models that are suitable for both the simulation and the control of the thermal, mechanical, and material properties of metal-based AM represents one of the key steps to reach the levels of quality and repeatability required by the industry. Indeed, such models are not only paramount to simulate and control the process but also to optimize its parameters and to predict the properties of the manufactured part as well as its compliance with the required specifications~\cite{RN120}.

One key component of a metal-based AM model concerns its capability to accurately represent the transient temperature field for a layer by layer process with moving laser heat source. Indeed, the transient temperature history is one of the main factors determining the thermal stress distribution and residual stress of the metallic part. As heat distributions are inherently modeled by Partial Differential Equations (PDEs), many finite element modeling approaches have been proposed in the literature for simulations. Three-dimensional convection models in the presence of driving forces for flow with moving laser beam were developed in~\cite{RN125,RN139}. These models essentially rely on the heat conduction equation which takes, for a workpiece of constant density $\rho$ and specific heat $C_p$, the following form:
\begin{equation*}
    \rho C_p \dfrac{\partial T}{\partial t} 
    = - \dfrac{\partial q}{\partial x} + Q
\end{equation*}
where $T$ is the temperature, $Q$ the heat source, and $q$ the heat flux vector which is described by the Fourier's conduction equation:
\begin{equation*}
    q = - k(T) \dfrac{\partial q}{\partial x}
\end{equation*}
with $k(T)$ the thermal conductivity. The heat source can be, for example, modeled by an ellipsoid~\cite{RN224}:
\begin{equation*}
    Q(t,x,y,z) = \dfrac{6 \sqrt{3} P \eta}{a b c \pi^{3/2}} \exp{\left( - \dfrac{3 x^2}{a^2} - \dfrac{3 y^2}{b^2} - \dfrac{3 (z+v_w t)^2}{c^2} \right)}
\end{equation*}
where $P$ is the incident laser power, $\eta$ the laser absorption efficiency, $a,b,c$ the parameters of the ellipsoid, and $v_w$ the laser velocity.
The boundary conditions of the PDE model the heat losses due to the convection $q_\mathrm{conv}$ expressed by Newton's law as:
\begin{equation*}
    q_\mathrm{conv} = h ( T_s - T_a )
\end{equation*}
where $T_s$ is the surface temperature, $T_a$ the ambiant temperature, $h$ the transfer coefficient; while the loss due to thermal radiations $q_\mathrm{rad}$ is described by Stephan-Boltzmann's law:
\begin{equation*}
    q_\mathrm{rad} = \epsilon \sigma (T_s^4 - T_a^4)
\end{equation*}
with $\sigma$ the Stephan-Boltzmann constant and $\epsilon$ the surface emissivity. Three-dimensional numerical models that take into account the multiple-line sintering under the heat source of a moving Gaussian laser beam were developed in~\cite{RN123,RN124,RN141}. To simulate the addition of layers with time, the method of element birth and death was proposed in~\cite{RN122}. A {3D} transient numerical model of multi-layer laser processes has been proposed in~\cite{RN202} to predict the geometry of the deposited material along with the temperature distribution and the thermal stress field. The complex interactions between the laser characteristics (e.g., power and scan speed) and the resulting geometry of the melting pool were modeled in~\cite{RN144}. Models for cladding and repair were proposed in~\cite{RN126}. In order to model the mechanical properties of the manufactured metallic part, thermo-mechanical models were developed~\cite{RN132}, allowing the analysis of thermal shape distortions~\cite{RN127,RN128,RN129,RN130,RN131,RN138}. Simulation strategies for modeling the material deposition in metal deposition processes were investigated in~\cite{RN133} using inactive or quiet elements methods. A critical review of the finite element methods for simulation of metallic powder bed AM processes is proposed in~\cite{RN140}.

Taking advantage of finite element methods, these PDE-based models can be used for the prediction of the AM processes (shape, mechanical properties, etc.). However, due to their inherent complexity (complex geometry of the part, laser-matter interactions, geometry of the melt pool, etc.), they are difficult to handle from a control system design viewpoint. Consequently, the development of finite dimensional reduced order models of the AM processes for control design is a research topic of primary interest. 

Black-box empirical models were developed in~\cite{RN147,RN148} for the control of the melt pool temperature and deposition height during a cladding process based on subspace model identification techniques. Specifically, the input-output behavior from the laser power to the melt pool temperature was identified experimentally as a second order (in~\cite{RN147}, fourth order in~\cite{RN148}) linear time-invariant model (in state space form). Similarly, a first-order transfer function was experimentally obtained in~\cite{RN149} for modeling the dynamics between the laser power and the width of the melt pool. In~\cite{RN155}, a second-order discrete model is identified for modeling the dynamics from the laser pulse energy to the height of the clad height. A first-order transfer function between the laser power and the melt pool temperature is identified based on graphical methods in~\cite{RN156}. In~\cite{RN157}, a first-order transfer function between a composite signal, mixing the laser scan speed, the laser power, and the powder flow rate, and the melt pool temperature is obtained based on a least-square method.  

Even if black-box empirical models can be successfully used for control design purposes, they suffer from several inherent limitations. One essential drawback of such black-box empirical models is that they require the application of the identification procedure to each of the operating points of interest (it includes ambient conditions, materials, required temperature, width, and depth of the melt pool, etc.). Furthermore, most of the black-box empirical models are fully linear while it is known that nonlinear effects occur in additive manufacturing processes (e.g., thermal radiation). In this context, there is a strong interest in the development of advanced physics-based models that are suitable for feedback control design. In~\cite{RN215,RN216}, a semi-empirical model was developed for the height control in laser solid freeform fabrication. Specifically, the dynamic response is empirically captured by a first-order dynamics while the steady-state value is obtained via physics-based consideration through a mass balance equation along with experimental identification procedures relying on least squares methods. A physics-based lumped-parameter model for metal deposition was developed in~\cite{RN217}. Derived based on physical considerations (such as the mass and energy balance of the melt pool puddle), the model describes the dynamics of both temperature and geometry of an ellipsoidal molten puddle. Such a model was further developed in~\cite{RN208,RN219,RN221} to layer-dependent process models and in~\cite{RN220} for improving the steady-state predictions of the melt pool geometrical characteristics. The later model consists of the following two sets of ODEs. The first one describes the mass conservation of the molten puddle:
\begin{equation*}
    \rho \dfrac{\mathrm{d} V}{\mathrm{d} t} + \rho A v = \mu f
\end{equation*}
and the second one describing the energy balance of the puddle:
\begin{equation*}
    \rho \dfrac{\mathrm{d} (V e)}{\mathrm{d} t} = - \rho A v e_b + P_S ,
\end{equation*}
where $\rho$ is the constant melt density, $\mu$ is the mass transfer efficiency, $V$ and $A$ are the volume and the maximal cross-sectional area of the molten puddle, respectively,  $f$ is the powder flow rate, $v$ is the laser scan speed, $e$ is the specific internal energy of the melt pool, $e_0$ is the specific energy of the solidified bead material, and $P_s$ is the total thermal transfer at the puddle surface.

\subsubsection{Low-level control strategies}

Laser cladding processes are governed by a large number of parameters and their respective interactions~\cite{RN148} (for example, laser, nozzle, powder and gas delivery system, substrate). Among the most important parameters to be controlled, temperature and the dimensions of the melt pool are perhaps the most important. These process parameters have a significant impact on the quality of the final product, and in particular on the dimensional accuracy. Control of the dimensional accuracy was identified in~\cite{RN196} as one of the ten most important challenges in AM. Geometrical inaccuracies can result from many factors~\cite{RN199}, including geometric errors introduced by the slicing process used to create the STL file~\cite{RN193,RN197} and thermal deformations resulting from stress gradient induced by the continuous cycle of rapid melting and solidification of the metal~\cite{RN194}. In order to avoid post-processing of the printed part with machine tools, which is both time-consuming and increases the production costs, prediction methods~\cite{RN204} and compensation strategies have been proposed in the design of the {3D} models. The latter consists of the integration of a shrinkage compensation aspect directly in the product design~\cite{RN195,RN198,RN206,RN207}.  It is in this context that feedback control of the cladding processes is paramount for ensuring the quality of the clad layers~\cite{RN222}. For instance, the benefit of the feedback control for the manufacturing of turbine blades has been experimentally highlighted in~\cite{RN147}. Indeed, while the uncontrolled manufacturing of the turbine blade induced inhomogeneities and geometrical inaccuracies the feedback control of both temperature and height of the melt pool significantly improved the quality of the manufactured part. The benefits of feedback control on microstructure, thermal distortion, and mechanical properties have also been illustrated in~\cite{RN212}. For this reason, many control strategies have been reported in the literature in order to ensure precise control of the temperature or the geometric characteristics of the melt pool. Even though some  design PDE model-based control strategies have been reported ~\cite{RN223}, the large majority of the documented approaches work with either black-box or (semi-empirical) lumped parameters models. In~\cite{RN149}, a camera-based feedback control strategy is proposed to control the width of the melt pool. The melt pool dimension are estimated in real time based on a greyscale camera via three main steps: Gaussian filtering to filter out isolated pixels resulting from hot particles; conversion of the image into a black and white image (using a threshold value corresponding to the boundary of the melt pool); and approximation of the pool-shape with an ellipse. The width of the melt pool is controlled by means of a PI controller. The control input is the laser power while the measurement is a filtered version (use of a second-order Butterworth filter) of the estimated width of the melt pool. In~\cite{RN148}, a dual-color pyrometer is used to sense the temperature of the melt pool. The control strategy consists of a predictive control algorithm with reference tracking of the temperature of the melt pool based on the adjustment of the laser power. In~\cite{RN201}, a PID controller is designed to regulate the temperature of the melt pool based on infrared image sensing (see also~\cite{RN203}). An opto-electronic sensor is also developed to monitor the powder flow rate, which is a paramount parameter for composite materials or alloys. In~\cite{RN150}, it is proposed to control both the powder flow rate and the area of the melt pool. The powder flow rate is controlled by direct feedback of the tracking error signal. The area of the melt pool is monitored in real time by an infrared image acquisition system. Then, image processing is carried out to evaluate the number of pixels inside the melt pool. This quantity is used as a feedback signal to control the melt pool area via a PID controller augmented with a feedforward to compensate for the existence of a time-lag in the control loop. Due to potential variations in the dynamics and melt pool characteristics as a function of the height and number of layers~\cite{RN208}, an iterative learning controller was designed in~\cite{RN211}. The feedback control of the clad height in laser solid freeform fabrication based on the development semi-empirical models was investigated in~\cite{RN215,RN216}. Different control strategies have been proposed and tested for the manufacturing of parts: in particular, feedforward PID control and sliding mode control. While offering different trade-offs in terms of response time and damping of the closed-loop system, these control strategies were successfully applied to the deposited layer in a nonplanar laser cladding process. In~\cite{RN157}, a composite control input, mixing the laser scan speed, the laser power, and the powder flow rate, are used to control the temperature of the melt pool. The controller strategy consists of a discrete time controller coupled with a Kalman Filter used to improve the closed-loop system performance by filtering the temperature signal provided by a sensor mounted on the nozzle.

Most of the aforementioned control strategies have been designed based on black-box empirical models. Taking advantage of the recent development of semi-empirical or physics-based lumped parameter models, model-based control strategies have also been recently reported. The authors in~\cite{RN217} investigate the control of width and height of the molten puddle for metal deposition by means of a PI controller coupled with a model-based estimator of the molten puddle geometrical characteristics. The feedback control of melt pool geometry and temperature in directed energy deposition has been investigated in~\cite{RN220}. The employed control strategy consists of a dynamic inversion ensuring the reference tracking of both temperature and height of the melt pool by using both laser power and scan speed as control inputs.

In order to improve both the performance and repeatability of cladding systems for industrial applications, it was proposed in~\cite{RN151} to take advantage of the flexibility of Field-Programmable Gate Arrays (FPGAs) for both monitoring and control purposes. The monitoring part consists of the measurement of the melt pool width~\cite{RN152,RN153}. In this setting, a CMOS camera is used to obtain a greyscale picture of the melt pool. Then, image processing is carried out by the FPGAs. This consists of three main steps: binarization via the selection of a gray level threshold; erosion to remove incorrectly saturated pixels; and width estimation~\cite{RN153}. Processing algorithms can also be employed to increase the robustness of monitoring~\cite{RN153}. Taking advantage of this FPGA-based monitoring, a PI controller has been developed in~\cite{RN154} for controlling the melt pool depth. Subsequent experimental results demonstrate a significant performance improvement of the FPGA-based solution when compared to existing PC-based solutions. In particular, as expected, the FPGA-based solution ensures a faster response time, yielding an increased quality of the cladding. In order to avoid a manual retuning of the controller gains depending on the current environmental configuration (environmental conditions, materials to be treated, geometry of the parts, characteristics of the laser, etc), the development of an adaptive controller with updating policy relying on fuzzy rules has been proposed in~\cite{RN151}.

\section{Creation and interaction with {3D} models}\label{sec: creation and iteration with 3D models}

Two of the key components in the {3D} printing workflow of functional parts are the design of the {3D} model of the part to be printed and the evaluation of its interaction with other existing parts. In this section, we discuss the limitations of the traditional {3D} approaches, as well as the potential remedies that have been developed in the literature.

\subsection{Limitations of traditional approaches}

We discuss first the limitations of the traditional approaches for {3D} modeling.

\subsubsection{Design of {3D} models}
One of the essential limitations of {3D} manufacturing is the strong gap existing between the tangible final product and its {3D} modeling. The outcome of the {3D} printing process is a physical object existing in the physical world. In contrast, the {3D} model used to print the object only exists in the virtual world, making it more difficult and less intuitive to appreciate the manner in which one can interact with it before printing. Advanced proprietary Computer-Aided Design (CAD) software, such as Catia or SolidWorks, can be used for {3D} modeling. They include numerous advanced features that enable the evaluation of the interactions between different parts, as well as finite element analyses for mechanical or thermal effects. However, even though these are useful features, much work remains to be done.  For example, CAD was originally developed for conventional manufacturing and does not embrace the specific needs of additive manufacturing design~\cite{RN196}. Most existing CAD-based software require a strong technical background and is not readily accessible to a large non-technical audience. While it is true that more accessible software for {3D} modeling have been developed and are freely available on the internet, these do not solve the inherent difficulty of testing human-machine interactions for manufacturing~\cite{RN4,RN60, RN40, RN5}, and it is  generally difficult for inexperienced people in {3D} modeling to correctly locate and place objects in a {3D} environment while only having access to a {2D} window into the {3D} space through a computer screen. Furthermore, the use of traditional mouse and keyboard are unintuitive for the user, and provide limited methods of interactions for assembling, creating, interacting, modifying, positioning, and shaping {3D} models within a three-dimensional environment. In this context, there is a clear need for developing new tools for interacting with {3D} models~\cite{RN87,RN53} as part of the design process.

\subsubsection{Interacting objects}
Most {3D} printed goods are stand-alone objects. In other words, the printed goods are essentially objects presenting very basic interactions with other goods. However, the most promising perspectives for AM rely on its capability to produce complex parts interacting with complex systems, including other {3D} printed parts, conventionally manufactured parts, mechanisms, electronics components, etc. A typical example of such a complex interaction is provided in the health field with prosthesis or dental implants. Each of these is unique because specifically tailored for a given person. Specifically, in these later examples, one has to ensure its full compatibility with the body of the person, from both integrity and functionality point of views.

One of the main difficulties in the design of interacting objects relies on the capability of the designer to {\em close-the-loop}. Indeed, {3D} modeling of stand-alone  objects is a relatively accessible task because it is an {\em open-loop} design. The workflow of the designer is straightforward, consisting mainly of creating a {3D} model, then sending it to the {3D} printer to get the final good. In the case of complex systems with {3D} printed parts interacting with other parts, the workflow is iterative as it requires the designer to close-the-loop. First, the designer must have a precise idea of the whole system and its different parts. Then, after designing a {3D} model of one of the parts, he must evaluate how it interacts with the other parts. Based on this evaluation, the designer modifies the initial {3D} model for improving its functionality. This loop is iterated multiple times until obtaining a suitable {3D} model of the part. The resulting model is then sent to the {3D} printer to get a physical realization. Physical tests and interactions experiments are then conducted to check its functionality. If the final result does not provide full satisfaction, a new design loop on the {3D} model is iterated.

\subsection{Advanced approaches}

With the technological advancements in the fields of virtual/augmented reality (VR/AR), image processing, haptics, and machine learning, extensive research has been carried out to develop interactive interfaces for providing more intuitive and realistic platforms for digital prototyping. In particular, much effort has been devoted to enable the consumers (novice users) to design their own customized products are discussed in this section.

\subsubsection{Sketch-based modeling}
The design of a {3D} model for the purpose of manufacturing requires technical familiarity with {3D} CAD tools. This requirement limits the involvement of end users (customers) in the design of the product for personal fabrication. Much effort has been made by the research community to develop interfaces for novice users that do not necessarily have engineering knowledge and skills. Sketch-based prototyping tools ~\cite{RN49} allow novice users to design linkage-based mechanisms for fast prototyping~\cite{RN82}. Similarly, an interactive system~\cite{RN35} that lets end users design toys (plushies) has been developed. Using this tool, the user can simulate the {3D} model of the toy by drawing the 2D sketch of its desired silhouette. A similar idea of coupling sketching with generative design tools (OptiStruct, solidThinking, Autodesk$^\mathrm{TM}$ Nastran Shape Generator, and Siemens$^\mathrm{TM}$ Frustum) to make the generative design more accessible to novice designers has been proposed in~\cite{RN92}. Although these interaction platforms allow users to achieve certain aesthetic goals while satisfying engineering constraints, they are unable to adapt designs according to a physical environment. AR and mixed reality provide more intuitive environments to bridge this gap between the physical and the digital worlds~\cite{RN54}. In particular, window-Shaping~\cite{RN91} is a design idea that integrates sketch/image-based {3D} modeling approach within a mixed reality interface and allows the user to design {3D} models on and around the physical objects.

\subsubsection{Gesture-based design}
Gesture-based interactions~\cite{RN20,RN21,RN36} require real-time hand tracking to recognize gestures. Existing solutions either use visual markers, haptic gloves or additional hardware for this purpose. FingARtips~\cite{RN10} presents a fingertip-based AR interface to interact with virtual objects. In this interface, the hand position is tracked using visual markers and vibrotactile actuators are used to provide haptic feedback. Tangible~3D~\cite{RN17} presents an immersive {3D} modeling system to create and interact with {3D} models using cameras and projectors. Similarly, situated modeling allows the user to create real-sized {3D} models using existing objects or an environment as a reference for physical guidance~\cite{RN32}. MirageTable~\cite{RN9}, an interactive system to merge real and virtual worlds, combines a depth camera, a curved screen, and a stereoscopic projector, to enable virtual {3D} model creation using gestures along with other interesting applications. Data Miming~\cite{RN14} uses an overhead camera to infer spatial objects from the user's gestures. Using a data miming approach, the user can describe the  physical objects with gestures and the interface then matches the input voxel representation of the gestures with the known {3D} model representation of a physical object. Similarly, another {3D} modeling system has been developed using an aerial imaging plate to project tablet PC screen mid-air and leap motion (motion sensor) to manipulate (move, scale and rotate) the virtual object using gestures to fit the physical object~\cite{RN46}. Although VR/AR interfaces are more intuitive for manipulating {3D} models they suffer from the fact that existing head-mounted displays (HMDs) are unable to provide precision due to mid-air gesture-based inputs. In these papers, gesture-based interfaces are, although intuitive, imprecise for generating {3D} models. In summary, the majority of tools-based on gesture-based interfaces, provide a platform to modify an existing {3D} model or to search from pre-existing {3D} models for non-expert users. To overcome some of these limitations, it was proposed in~\cite{RN42} to switch between classic CAD interface on a monitor (precise input but counter-intuitive) and AR mode (imprecise but intuitive) for designing {3D} models.

Interaction with virtual objects can be counter-intuitive relying solely on visual cues and gestures. Use of tangible tools (such as haptic gloves or additional hardware) for the creation or modification of virtual models makes this experience more intuitive. Surface drawing~\cite{RN22}, a semi-immersive virtual environment, in addition to hand tracking for virtual {3D} strokes, uses physical tools like tongs, as well as erasers and magnetic tools, for shape refinement of the {3D} model. Twister~\cite{RN18} is a manipulation tool that uses a 6 DoF magnetic tracker in each hand and allows the user to create or modify (tilt, twist or bend) {3D} shapes using both hands simultaneously. Digits~\cite{RN22} is a wrist-worn sensor that estimates the {3D} pose of the user's hand, hence enabling natural hand manipulations in the digital domain without requiring depth cameras or data gloves. Based on interactive situated AR systems like HoloDesk~\cite{RN13} and Holo Tabletop~\cite{RN45}, MixFab~\cite{RN1} is an immersive augmented reality environment that lowers the barrier for non-professional designers to engage in personal fabrication. This fabrication system allows the user to sketch and extrude the virtual artifact using hands. The hand gestures are recognized through a single depth camera whereas a motorized turntable is installed for {3D} scanning of the physical object. A user can use this digital model of the scanned physical artifact as a size or shape reference to integrate the physical object in the design process.         

\subsubsection{Scan-based modeling}

Now we discuss the possibility to use objects from the real world for either {3D} modeling or the evaluation of the interactivity of the {3D} with objects from the physical world.

\paragraph{3D modeling}
AR has proven to offer significant potential in providing platforms for rapid prototyping of complex systems. The dynamics of interaction between different components of such systems can be simulated to support iterative design prior to manufacturing. A similar approach has been used by combining paper craft, augmented reality, and virtual simulation for rapid prototyping and experimentation of complex systems, such as bicycle gears and the human circulatory system~\cite{RN25}. Scan-based interfaces allow the user to design digital models based on physical artifacts. For example, Tactum~\cite{RN38}, enables non-expert users to design products for their body (forearm) by scanning the forearm through a depth camera (Kinect) and then  using skin as an interactive input surface for designing {3D} models according to the scanned body part. Similarly, with RealFusion~\cite{RN43}, an interactive workflow allows novice designers to express their creative ideas through the manipulation of the digital models of the real objects. This interface allows the user to scan physical objects using a depth sensor and lets the user modify the scanned digital model utilizing mid-air interactions with a smartphone. 

\paragraph{Interaction with existing artifacts}
While {3D} modeling of stand-alone objects is a relatively accessible task, the design of fully functional artifacts interacting with other parts, such as hinges, remains challenging. This kind of fabrication can be achieved either by post-assembly or by directly integrating electrical components into {3D} printed objects. The post-assembly fabrication process includes printing the housing for electrical components, that can later be adapted to an existing physical artifact. Printy~\cite{RN90} is an augmented fabrication system that allows a novice user to design a customized {3D} model based on a modular circuit and add-on (i.e. a button) description to modify an existing {3D} model. Using a web-based interface, the user can also add cloud-based interactivity to their custom designs.  ModelCraft~\cite{RN79} is a syntax-based approach that can be used as a plug-in for SolidWorks. Starting from a {3D} model in SolidWorks, this system generates a {2D} pattern by unfolding the {3D} model. The {2D} pattern is then printed on a paper which is cut and assembled as a paper-based {3D} model. Using a Logitech io2$^\mathrm{TM}$ pen, the user can annotate and edit the paper-based {3D} model. The CAD model then renders these annotations to edit the digital model accordingly. Inspired by ModelCraft, Makers' Marks~\cite{RN39} is another annotation-based system to fabricate functional objects by combining existing parts with custom-designed enclosures. Users can use this system to scan their sculpture made from clay and annotated with stickers to show the placement of the functional components (e.g., mechanical hinges, electronic components). Makers' Marks then creates {3D} geometry of the sculpture and replaces the annotations with existing {3D} models of the functional components. Another technique to add existing objects within {3D} printed parts is to the design of internal pipes and cavities within {3D} models through path planning. The pipes then can be inserted with media post-print, and then these pipes can be used to either enable input or display outputs such as illuminations or some forms of haptic feedback~\cite{RN37}. RetroFab~\cite{RN41} is another augmented fabrication environment that enables a non-expert user to retrofit physical interfaces. RetroFab generates a {3D} model of an existing object through scanning and then enables the user to design and place various electronic components (actuators and sensors). The housing for these electronic components is automatically generated to fit with the existing physical object. These kinds of systems enable consumers to design their own control interfaces for household items and print customized enclosures to house the required electronic components. Capricate~\cite{RN57} is a fabrication pipeline that allows the user to design and print capacitive touch sensors embedded in {3D} printed objects instead of using a post-assembly approach to add interactive capabilities in {3D} objects.

\subsubsection{Tools-based design}

In order to make the interaction with the virtual environment more intuitive, works have explored the use of instrumented tools. In this section we give an overview of this work.

\paragraph{Tools-based {3D} modeling}
In an effort to make the traditional tooling experience applicable for crafting virtual models, researchers have focused on developing tools that can be used for handcrafting digital models. For example, a mixed reality handcrafting system has been developed in~\cite{RN3} using physical devices to imitate tools (knife, hammer, tweezers). These physical tools can be used to perform move, cut and join operations, on {3D} models in an intuitive manner. These physical tools are equipped to provide tactile sensations, and operational feedback through vibrations and sounds to make the experience more realistic~\cite{RN24}. Tools can also be used to involve, in a direct manner, the user in the manufacturing process. For instance, it was reported in~\cite{RN19} the possibility to draft directly on the workpiece by means of a hand-held laser pointer. The system tracks the pointer, generates a clean path, and cuts accordingly the workpiece using a laser cutter. 

Specifying dimensions and angles can be a tedious procedure for traditional computer-aided design tools. As the designed model will, once printed, interact with real-world artifacts, it is generally necessary to measure certain physical characteristics of existing objects. Then, this measurement information is entered into the software through a computer screen, a mouse, and a keyboard. Such a procedure introduces a gap between the measurement in the real world and its impact on the virtual {3D} model. To narrow this gap, it has been proposed to resort to a digitized version of traditional measurement tools (e.g., calipers, protractors) allowing a bidirectional transfer between the real world and the virtual environment~\cite{RN26}. This enables the specification of the physical features of a {3D} model in an intuitive manner based on direct measurements on physical objects. Second, the presence of actuators in the tools enable the physical representation of physical dimensions specified directly in the virtual environment.

Another approach was proposed in~\cite{RN47} for the creation of containers for fitting real objects in a virtual environment. Instead of measuring size, via augmented tools, the proposed approach consists first of capturing a picture of the object. Then this image is projected on the build plate of the printer. Finally, by means of a tactile screen, the user can draw directly on the build plate to either create or modify an existing model by taking advantage of the captured image.

The use of building blocks (e.g., lego bricks) for fast {3D} modeling of functional objects has been investigated in\cite{RN8}. Specifically, it has been proposed to use instrumented and sensed construction toys (physical building blocks), combining embedded computation, vision-based acquisition, and graphical interpretation, for easy-to-use and tangible {3D} modeling. Similar approaches for fast prototyping and building of structures were reported in~\cite{RN51} by assembling struts- and hubs-primitives, and in~\cite{RN85} based on cubes. 

Finally, the design of complex {3D} geometries exhibiting many degrees of freedom can be a difficult task with conventional hardware inputs such as a keyboard and a mouse. To solve this problem, it was proposed in~\cite{RN58,RN88} to use a shape-sensing strip for capturing curves of surfaces that exhibit complex geometries. In this setting, the user deforms, directly with his hands and in the real world, the shape sensing-strip. Then, the associated geometry is captured by means of a linear array of strain gauges located along the strip. The same type of approach has been developed in~\cite{RN89} for the articulation of {3D} characters, by means of the deformation by hand, of skeletal trees.

\paragraph{3D model-based assisted manufacturing}
Most research projects on tool-based design for advanced manufacturing strive to develop tools to help the user in the design of {3D} models. Certain projects also promote the emergence of a new fabrication approach combining digital fabrication and craft. For instance, the use of building blocks (e.g., lego bricks) for rapid prototyping of functional objects such as a head-mounted display or soap holder has been investigated in~\cite{RN81}. The developed approach substitutes parts of {3D} models with building blocks, while the user can specify the parts of the model that need to be printed. Such a mix of building blocks and {3D} printed parts speeds up the fabrication process and is thus suited for fast prototyping. In another example, a freehand digital sculpting tool developed in the framework of subtractive manufacturing, was reported in~\cite{RN28,RN29,RN30}. The system consists of a milling device monitored by a computer. Based on a pre-defined {3D} model, the computer allows sculpting, except when the milling reaches the surface of the {3D} model. Similarly, a mixed reality environment was developed in~\cite{RN56} for the drawing of {3D} wire structures by means of a {3D} extruder pen. Here the mixed reality environment is used as a guide for the user by superimposing onto the drawn structure a projection of the {3D} model.

\subsubsection{Haptic Interactions}
Haptic interfaces are devices that generate mechanical signals to stimulate kinesthetic and/or tactile senses of the human. These devices aim at providing force feedback for improving the interactivity within a virtual environment~\cite{RN68}. Unlike traditional interfaces that take advantage of visual and auditory senses for interactions in the virtual world, haptic interfaces allow the user to use the sense of touch to perceive rich and detailed information about the virtual object. In the context of this work, haptic feedback can be categorized as tactile feedback and kinesthetic feedback.

Tactile (cutaneous) feedback is related to sensing the pressure on the skin surface. The patterns of these sensations, perceived through the biological receptors spread across the whole body, are interpreted by the brain as weight, size, and texture of an object. Vibrotactile feedback is the most traditional and frequently used tactile feedback integrated in our mobile phones and game controllers. These types of actuators are fairly limited in conveying the shape, size, and texture of an object. Therefore, haptic devices are required which can offer more than buzzing and rumbling to the hand. Human fingertips are quite sensitive skin areas to sense a surface smoothness or texture. Therefore, attempts have been made to use actuators for fingertips to enable tactile feedback~\cite{RN76,RN69}. In these types of haptic interfaces, miniaturized DC motors with cables or belts are placed on the nails to generate controlled pressure on the fingertip to render a weight perception. NormalTouch and TextureTouch~\cite{RN72} present a mechanically actuated handheld controller for haptic shape rendering. This controller consists of a tiltable and extrudable platform for the finger to render the virtual object surface. To render fine-grained surface texture details, a $4 \times 4$  matrix  of  actuated pins are placed underneath the user's fingertip. To enable the user to touch the physical objects along with virtual objects, nail mounted tactile feedback has been proposed~\cite{RN62}. This device contains a voice coil and tactile sensations are produced by controlling the modulation of waveforms exciting the coil. The use of ultrasonic actuators embedded into head-mounted displays has also been proposed for tactile feedback in~\cite{RN75}. However, this kind of interface can only be used for VR headsets.  

Kinesthetic feedback is related to the feedback gathered from the sensors embedded in muscles, tendons, and joints. This type of feedback is used to perceive size, weight, and position of the object relative to the body. Kinesthetic feedback interfaces prevent the user hands or body from penetrating through the virtual object and hence, provide a more realistic experience. Exoskeleton glove-based interfaces take advantage of this feedback~\cite{RN61}. Electrical muscle stimulation (EMS) is another type of interface that is based on kinesthetic feedback that has been explored to make mixed reality experience more realistic ~\cite{RN2,RN63,RN73}.

Haptic feedback technologies also offer promising approaches for the improvement of AR experiences in manufacturing. For example, the use of a haptic feedback input device for navigating in CAD environments was reported in~\cite{RN50}. In this setting, the user manipulates the camera in the 3D environment by means of a tangible tool providing force feedback when a virtual obstacle is encountered. In~\cite{RN70}, the use of different feedback methods such as visual, pressure-based tactile, and vibrotactile feedbacks, have been investigated for improving human-machine interactions. Although haptic feedback-based direct interactions appear to be less robust and slower than indirect controller-based interactions, the former greatly improve both functionality and ergonomics in the manipulation of virtual objects. Among the great variety of applications, one can find a smartwatch with force feedback~\cite{RN64,RN65} and haptic feedback in robot-mediated surgery~\cite{RN66} also incorporating thermal feedback~\cite{RN71}.

\subsubsection{Toward an integrated virtual design and physical shaping}
Conventional additive manufacturing is a unidirectional process. First, a {3D} model of the part is built in the digital world. Then the part is manufactured. Such a workflow presents two fundamental limitations. First, it does not allow an iterative design in the sense that a printed artifact cannot be modified; any modification requires the printing of a new version from scratch. Second, while any modification on the original {3D} model will impact the final printed object, reshaping the physical object will have no influence on the virtual model. In this context, attempts have been made to provide more flexible design processes enabling iterative design and bidirectional interactions between virtual models and printed goods. An iterative technique allowing the patching of existing objects was presented in~\cite{RN93}. In this setting, the already printed object is mounted into the 3D printer while both original and modified CAD models are used to generate the tool trajectory to patch the object. The
problem of synchronizing the CAD model and the physical model has been investigated in~\cite{RN27,RN44}. After completing a 3D scanning of the physical object, an algorithm is used to detect the changes (either additive or subtractive) and then the associated {3D} model is updated. Another approach aimed at detecting touch and its characteristics (position on the object and applied force) for increasing the interactivity between the {3D} model and the printed good was reported in~\cite{RN59}. Such integrated virtual designs and physical shaping was further developed in~\cite{RN52} by using a robotic modeling assistant (RoMA) for simultaneous {3D} modeling and {3D} printing based on augmented reality. By merging the {3D} modeling environment with the printing workspace, RoMA enables the user to create a {3D} model directly within the printing workspace. In this setting, the partially printed object can be used as a tangible reference for the design of new elements. The idea of direct interactions between real and virtual worlds within the printing workspace has also been developed in several reported works. For example, the possibility to superimpose a hologram of the {3D} model on top of the currently printed object has been reported in~\cite{RN48}. This setup allows the user to design CAD models that are directly projected in the 3D printer workspace in real scale. Such an approach was also developed in~\cite{RN77} with application to the real-time monitoring of the geometrical accuracy of the printed object. In~\cite{RN230}, a layer-by-layer 3D model reconstruction, using a novel scan-based method, for the real-time monitoring of additive manufacturing processes is proposed. This method enables the user, directly during the printing process, to view and detect potential defects, not only at the surface but also in the inner layers of the printed object using an AR interface.

\subsubsection{Complementary approaches}
In this subsection, we discuss approaches that are emerging in this rapidly changing field. For example, inspired by clay modeling, a digital clay interface allowing the impression of shapes from physical objects into digital models by deforming a malleable gel input device was reported in~\cite{RN12}. In~\cite{RN23}, also inspired by clay modeling, the user sculpts virtual models by manipulating physical prop and an annotations-based system can be used for integrating complex components (hinges, electronics)~\cite{RN39}. While scanning techniques are commonly used for the {3D} modeling of real objects, the possibility to use {2D} inputs (pictures) has also been investigated in~\cite{RN34,RN55}. In the near future, the emergence of shape-changing interfaces~\cite{RN84} as a new method for interacting with computers could be a valuable technology for {3D} modeling. The recent developments in machine learning for the improvement of human-machine interactions also offer opportunities for AM~\cite{RN7}.

\section{Concluding remarks}\label{sec: perspectives and conclusion}

Our objective in this paper was to review the application of feedback ideas in the area of additive manufacturing. Both the application of feedback control to the 3D printing process, and the application of feedback theory to enable users to interact better with machines, are reviewed. We believe that this paper is first such detailed review presented in the literature. Our future work will build on these ideas to enable real interaction real and virtual worlds as part of a scaled hardware-in-the-loop platform.

%

\bibliographystyle{IEEEtran}
\bibliography{IEEEabrv,Library_human_machine_interaction}

\end{document}